\documentstyle[fleqn,epsfig]{article}  
\newcommand {\delm}{$\Delta m^{2}$\hspace{1mm}} 
\newcommand {\sqsin}{$\sin^{2} 2\theta$\hspace{1mm}}
\newcommand {\delmns}{$\Delta m^{2}$} 
\newcommand {\sqsinns}{$\sin^{2} 2\theta$}
\newcommand {\sqdm}{$\sin^{2} 2\theta,\Delta m^{2}$\hspace{1mm}}
\newcommand {\sqdmns}{$\sin^{2} 2\theta,\Delta m^{2}$} 
\newcommand {\loe}{$L/E$\hspace{1mm}} 
\newcommand {\ltenloe}{$\log_{10}L/E$\hspace{1mm}}
\newcommand {\numutonutau}{$\nu_\mu \rightarrow \nu_\tau$\hspace{1mm}}
\newcommand {\muflvr}{\mbox{$\mu$-flavor}\hspace{1mm}}
\newcommand {\eflvr}{\mbox{e-flavor}\hspace{1mm}}
\newcommand {\muflvrns}{\mbox{$\mu$-flavor}}
\newcommand {\eflvrns}{\mbox{e-flavor}}
\newcommand{\numu}{\mbox{$\nu_{\mu}$}\hspace{1mm}}                   
\newcommand{\nue}{\mbox{$\nu_{e}$}\hspace{1mm}}                      
\newcommand{\nus}{\mbox{$\nu_{s}$}\hspace{1mm}}                      
\newcommand{\nutau}{\mbox{$\nu_{\tau}$}\hspace{1mm}}                 

\title{   
{\bf Observation of Atmospheric Neutrino Oscillations \\ 
in Soudan 2} 
\vskip 30pt 
} 

\author{ 
 M.~Sanchez$^5$, W.W.M.~Allison$^3$, G.J.~Alner$^4$, D.S.~Ayres$^1$, W.L.~Barrett$^6$,\\ 
P.M.~Border$^2$, J.H.~Cobb$^3$, D.J.A.~Cockerill$^4$, H.~Courant$^2$,\\
 D.M.~Demuth$^2$,
T.H.~Fields$^1$, H.R.~Gallagher$^5$, M.C.~Goodman$^1$,  \\ 
T.~Joffe-Minor$^1$, T.~Kafka$^5$, S.M.S.~Kasahara$^2$, 
P.J.~Litchfield$^2$, \\ W.A.~Mann$^5$, M.L.~Marshak$^2$, 
R.H.~Milburn$^5$, W.H.~Miller$^2$,  L.~Mualem$^2$, \\ J.K.~Nelson$^2$, 
A.~Napier$^5$, W.P.~Oliver$^5$, G.F.~Pearce$^4$, E.A.~Peterson$^2$,\\ 
D.A.~Petyt$^2$, K.~Ruddick$^2$, J.~Schneps$^5$, 
\\ A.~Sousa$^5$, B.~Speakman$^2$, J.L.~Thron$^1$, N.~West$^3$\\ 
\\ 
$^1${\it Argonne National Laboratory, Argonne, IL 60439}\\ 
$^2${\it University of Minnesota, Minneapolis, MN 55455}\\ 
$^3${\it Department of Physics, University of Oxford, Oxford OX1 3RH, UK}\\ 
$^4${\it Rutherford Appleton Laboratory, Chilton, Didcot, Oxfordshire 
 OX11 0QX, UK}\\ 
$^5${\it Tufts University, Medford, MA 02155}\\ 
$^6${\it Western Washington University, Bellingham, WA 98225}\\ 
}

\begin{document}

\maketitle 

\thispagestyle{empty} 

\begin{abstract} 
\normalsize 

The effects of oscillations of atmospheric \numu are observed in the  5.90 fiducial kiloton-year exposure of the Soudan~2 detector. An unbinned maximum likelihood analysis of the neutrino $L/E$ distribution has been carried out using the Feldman-Cousins prescription. The probability of the no oscillation  hypothesis  is $5.8 \times 10^{-4}$. The 90\% confidence allowed region in the \sqdm plane is presented.

\vskip 20pt 
\noindent PACS numbers: 14.60.Lm, 14.60.Pq, 95.85.Ry 

\end{abstract} 
\eject 
\large

\section{Introduction} 

     The discovery of neutrino oscillations in both atmospheric and solar neutrinos and thus the establishment of neutrino mass has been one of the major advances in particle physics of the past decade. The evidence generally regarded as  establishing neutrino oscillations was the observation by the Super-Kamiokande collaboration of a variation of the atmospheric neutrino event rate with zenith angle \cite{Super-K_OSC}. Solar neutrino oscillations have been confirmed by the SNO experiment's measurement of the neutral current event rate \cite{SNO} and the detection of the disappearance of reactor neutrinos by KamLAND \cite{KamLAND}.  The K2K experiment has provided supporting evidence for \numu oscillations \cite{K2K} but there has been no detailed confirmation of the Super-K effect in atmospheric neutrinos.  Observations of a deficit of atmospheric \numu have been published by this experiment \cite{Allison2}, and by Kamiokande \cite{Kam}, IMB \cite{IMB} and MACRO \cite{MACRO_OSC}.  The analysis reported here is the first independent confirmation of atmospheric neutrino oscillations using fully reconstructed neutrino interactions and covering the complete range of zenith angles.
   
  The data used are from the 5.90 fiducial kton-year exposure of the Soudan~2 detector. Soudan~2 was originally designed to study proton decay and thus has excellent resolution and pattern recognition properties in the visible energy region around 1 GeV where the peak in the atmospheric neutrino event rate occurs.  Although the exposure of the experiment is less than that of Super-K, the full event reconstruction, and thus good energy and direction resolution for the incident neutrino, compensates to some extent for the smaller number of events.

  The data are analyzed using an unbinned likelihood method based on the Feldman-Cousins prescription \cite{Feldman_Cousins}.  The probability of the no oscillation hypothesis is $ 5.8 \times 10^{-4}$.  The 90\% confidence allowed region in the \sqdm plane is determined and is consistent with that published by Super-K. 

\section{Detector and Data Exposure} 

\begin{figure}[htb]
\centerline{\epsfig{file={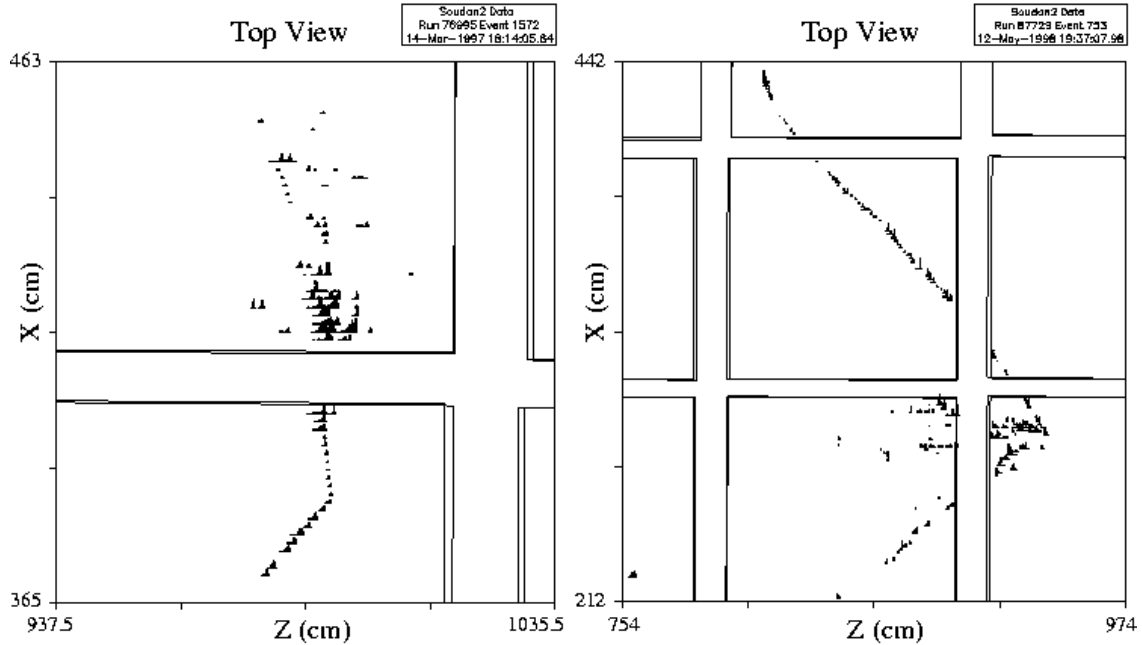} ,width=15.0cm}} 
\caption{Two neutrino interactions in Soudan 2.  The event on the left is a quasi-elastic \nue interaction producing a proton and an electron.  The electron travels about one radiation length before showering.  The proton is easily recognizable by its heavy ionization (large symbols) and its lack of Coulomb scattering.  The event on the right has a long non-interacting muon track, which shows typical Coulomb scattering, and a hadronic shower at the vertex. The shower contains a charged pion and at least two electromagnetic showers.}
\label{fig:events}
\end{figure}

   Soudan 2 is a 963 metric ton (770 tons fiducial) 
iron tracking calorimeter with a honeycomb geometry 
which operates as a time projection chamber. 
The detector is located at a depth of 2070 meters--water--equivalent 
on the 27th level of the Soudan Underground Mine State Park 
in northern Minnesota.  The  calorimeter 
started  data taking  
in April 1989 and ceased operation in June 2001 by which time a total exposure of 7.36 kton-years, corresponding to a fiducial exposure of 5.90 kton-years, had been obtained.

  The calorimeter's active elements are 1 m long, 1.5 cm diameter hytrel plastic drift tubes filled with an argon-CO$_2$ gas mixture. The tubes are encased in a honeycomb matrix of 1.6~mm thick corrugated steel plates.  Electrons deposited in the gas by the passage of charged particles drifted to the tube ends under the influence of an electric field.  At the tube ends the electrons were amplified by vertical anode wires which read out a column of tubes.  A horizontal cathode strip read out the induced charge and the third coordinate was provided by the drift time.  The ionization deposited was measured by the anode pulse height.   
   The steel sheets are stacked to form 
1$\times$1$\times$2.5 m$^3$, 4.3 ton modules 
from which the calorimeter was assembled 
in building-block fashion.  More details of the construction of the detector and its properties can be found in Ref.~\cite{S2:NIM_A376_A381}. 

Surrounding the tracking calorimeter on all sides but mounted on the 
cavern walls, well separated from the outer surfaces of the calorimeter, is 
a 1700 m$^{2}$ active shield array of two or three layers of 
proportional tubes~\cite{S2:NIM_A276}.  The shield  
tagged the presence of cosmic ray muons in time with events in the main calorimeter and thus identified background events, either produced directly by the muons or initiated by secondary particles coming from muon interactions in the rock walls of the cavern. 
   
   Calibration of the calorimeter response 
was carried out at the Rutherford Laboratory ISIS spallation 
neutron facility using test beams of  
pions, electrons, muons, and protons~\cite{Wall_PRD62_00}. 
Spatial resolutions for track reconstruction and for vertex 
placement in anode, cathode, and drift time coordinates are of the 
same scale as the drift tube radii, $\approx$ 0.7 cm. 

   Soudan 2 has several advantages over water Cherenkov detectors.  Images approaching the quality of bubble chamber events are obtained.   Ionizing particles having non-relativistic as well 
as relativistic momenta are detected via their energy loss in the gas.
Protons are readily distinguished from $\pi^{\pm}$ and $\mu^{\pm}$ 
tracks by their ionization and lack of multiple Coulomb scattering.  Muons from \numu charged current (CC) 
reactions are prompt tracks without 
secondary scatters. Prompt e$^{\pm}$ showers from \nue CC reactions 
are distinguished from photon showers on the basis of their proximity 
to the primary vertex. Since Soudan~2 has no magnetic field and thus only limited charge identification,  $\nu$ and $\bar{\nu}$ reactions are not separated.

Two examples of the event definition provided by the detector are shown in Fig.~\ref{fig:events}.  The event on the left is a quasi-elastic \nue interaction producing a short proton and an electron which travels approximately one radiation length before showering.  The event on the right is an inelastic \numu interaction.  The long non-interacting muon track is accompanied by a hadronic shower, including a charged pion and at least two gamma showers.  

  The excellent imaging and particle identification  offer good 
 determination of the energy and direction of the incident neutrino and thus the path length from its production point in the atmosphere.  Especially advantageous is  the reconstruction of 
 quasi-elastic reactions, where the recoil proton is observed with approximately 40\% efficiency,  and
 complicated multiprong topologies. The correlation of the outgoing lepton direction and energy with the incident neutrino direction and energy is poor at low energies. Improvements in the resolution of neutrino path length divided by energy, \loe, by factors of 2 and 3 are readily obtained by reconstructing both the lepton and the hadronic final state.

\section{ Event Classes and Processing}
\label{sec:datasamples}

\subsection{Containment classes}
  Events are divided into two containment classes. 
\begin{enumerate}
\item {\it Events that are fully contained within the detector (FCE)}.  Containment is defined by the requirement that no portion of the event approaches closer than 20 cm to the exterior of the detector and that no particle in the event could enter or escape the detector through the space between modules.  The containment criterion limits high energy \numu events to those with a muon of energy less than around 1 GeV.
\item {\it Events that are partially contained, in which only the produced lepton exits the detector (PCE)}.  These events recover a fraction of the high energy \numu events rejected by the containment criterion. As the muon does not stop in the detector, its energy from range cannot be measured.  Instead an estimate of the energy was obtained from the observed range, with a small added correction based on the amount of multiple scattering on the track. Monte Carlo (MC) studies showed that the effect of the poorer energy measurement on the \loe resolution was small.  Since the timing resolution of Soudan~2 is insufficient to determine the $\mu$ direction, a stopping, downward-going, cosmic ray muon could mimic an upward-going PCE having little or no hadronic vertex activity. Up-down asymmetric cuts on the exiting track and the event vertex properties are required to reduce this contamination to negligible proportions \cite{PCE_Petyt}. 
\end{enumerate}

\subsection{Veto Shield classes}
\label{sec:evclass}
     Two classes of events are defined on the basis of the presence or absence of hits in the veto shield.
\begin{enumerate}
\item {\it Quiet shield events}. These events have no in-time hits in the veto shield except for those associated with the leaving lepton in PCEs. They are candidates for neutrino interactions but also contain a small background of events produced by cosmic ray muons.  They will be called ``{\it qs-data}'' events.
\item {\it Shield tagged events}. Events initiated by the passage of a cosmic ray muon generally have in-time hits in the veto shield. The hits may be caused by the muon itself or by secondary charged particles from  the interaction of the muon in the rock surrounding the detector. Secondary  neutral particles can enter the detector and interact, mimicking neutrino interactions.  Events flagged with in-time veto shield hits will be called  ``{\it rock}'' events.
\end{enumerate}  

The average shield efficiency for detection of a minimum ionizing particle was measured to be 94\%. Study of events with a single shield hit showed that the contamination of qs-data events  by cosmic ray muons which pass through the shield and enter the detector is negligible.  It was however possible for neutrons and gamma rays to enter the detector with no identifying shield hit when all of the charged particles associated with the production event in the rock passed outside the shield or were not detected due to shield inefficiency.  These quiet shield rock events (called ``{\it qs-rock}'') are a background to the neutrino sample. They may be statistically distinguished from neutrino events by the depth distribution of the interaction vertices, as described in Sect.~\ref{dataa}.

\subsection{Topology classes}
The background from qs-rock events is significantly different in low and high multiplicity events and in low multiplicity electron and muon samples. The FCE data are thus further divided into topology classes.
\begin{enumerate}
\item  {\it Events with a single track-like particle with or without a recoil proton, called ``tracks''}.  These are mostly quasi-elastic \numu interactions and are assigned ``{\it $\mu$-flavor}''.
\item  {\it Events with a single showering particle with or without a recoil proton called ``showers''}.  These are mostly quasi-elastic \nue interactions and are assigned ``{\it e-flavor}''.
\item {\it Events with multiple outgoing tracks and/or showers called ``multiprongs''}.  These can be of either flavor.  The flavor is assigned according to whether the highest energy secondary is a non-scattering track (\muflvrns) or shower (\eflvrns). A small fraction of multiprongs have no muon or electron candidate and are defined as neutral current, {\it``NC''}.  A further small sample had no obvious flavor and are  defined as ambiguous.  The NC events are too few to provide constraints on the oscillation analysis but they and the ambiguous events are added into the event total, contributing to  the flux normalization.  Through the rest of the paper, unless otherwise specified, the heading NC includes both neutral current and ambiguous events.
\end{enumerate}

\subsection{High resolution sample}

 At low neutrino energies the correlation between the direction of the outgoing lepton  and the  incoming neutrino is poor. The ability of  Soudan~2 to reconstruct the recoil proton from  quasi-elastic interactions and the low energy particles from inelastic reactions gives a major improvement in the neutrino pointing and energy  resolution. To take advantage of this, the events are finally divided into two samples depending on  the \loe resolution.
\begin{enumerate}
\item {\it A high resolution ``HiRes'' sample}:
\begin{enumerate}
\item  events with a single lepton of kinetic energy $>$600 MeV,
\item  events with a single lepton of kinetic energy $>$150 MeV with a reconstructed recoil proton,
\item  multiprong events with lepton kinetic energy $>$250 MeV, total visible momentum $>$450 MeV/$c$ and total visible energy $>$700 MeV,
\item partially contained events.
\end{enumerate}
The mean neutrino pointing error for events in this sample is $33^\circ$ for \numu FCE, $21^\circ$ for \nue FCE and $14^\circ$ for \numu PCE.  This yields a mean error in \ltenloe of approximately 0.2.
\item {\it A low resolution ``LoRes'' sample}, comprising all other events.
\end{enumerate} 

\subsection{Monte Carlo neutrino events} 
\label{sec:mclass} 
  To avoid biases due to the scanning described in Sect.~\ref{datap} and to provide a blind analysis, Monte Carlo events were inserted into the data stream as the data were taken. A Monte Carlo sample 6.1 times the expected data sample in this exposure, assuming no oscillations, was included.    The Monte Carlo and data events were then processed simultaneously and identically through the analysis chain.  The Monte Carlo representation of the detector and background noise was of sufficient quality that a human scanner could not distinguish between data and Monte Carlo events.  The event generation was carried out using the NEUGEN package \cite{neugen} and the particle transport using GHEISHA \cite{geisha} and EGS \cite{egs}.  The incident neutrino flux used in the generation of events was that provided by the Bartol group at the start of the experiment \cite{bartol89}.  Later calculations of the neutrino flux were accommodated by weighting the generated events. The main analysis described below used the Bartol 96 flux \cite{bartol96}.  A recent three-dimensional calculation of Battistoni et al. \cite{Battistoni_3D} has also been used. The variation in the flux during the solar cycle was taken into account using  measured neutron monitor data for normalization \cite{nmon}.  The Monte Carlo events were superimposed on random triggers which represent the background noise in the detector, mostly low energy gammas from radioactive decays or electronic noise.  This simulated the effect of random noise on the event recognition and reconstruction and the random vetoing of events (4.8\%) due to noise in the veto shield.

\subsection{Data processing}
\label{datap}

\begin{figure}[htb]
\centerline{\epsfig{file={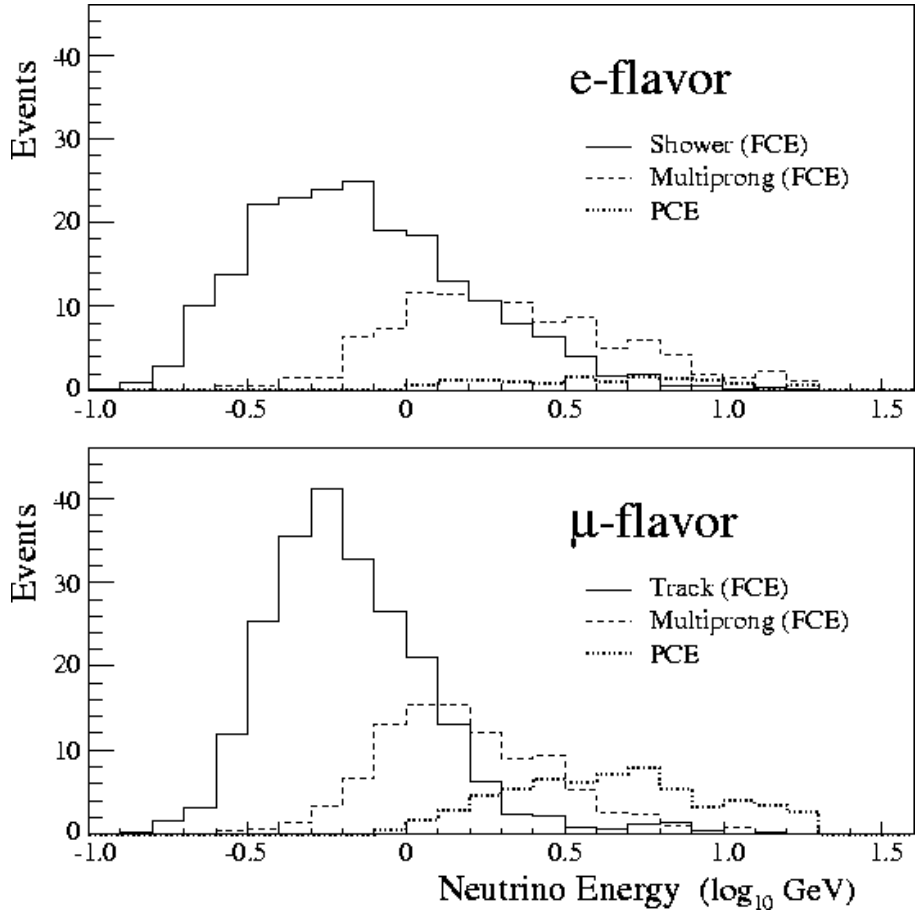} ,width=12.0cm}}  
\caption{Unoscillated Monte Carlo neutrino energy distributions for the fully contained shower or track, fully contained multiprong and PCE classes.  The top plot is for \eflvr events  and the bottom plot for \muflvr events.}
\label{fig:energy}
\end{figure}

      The detector trigger rate was approximately 0.5 Hz.  About half of the triggered events were cosmic ray muons and half electronic noise or the sum of random low energy gamma radiation.  The trigger required 7 anodes or 8 cathodes to have signals in any contiguous set of 16 multiplexed channels.  The  50\% efficiency thresholds are approximately 100 MeV for single electrons and 150 MeV kinetic energy for single muons.

      The data, including the Monte Carlo events, were processed through the standard Soudan~2 reconstruction code to select possible contained and partially contained events.  Approximately 0.1\% of events were retained for further analysis.  These events were double scanned to verify containment and remove remaining noise backgrounds.  Surviving events were checked and assigned flavor by three independent physicist scans. Throughout this process qs-data, rock and MC events were treated identically, without their origin being known to the scanners. The classifications based on confinement, topology and resolution were also applied equally to the three data types. 

Accepted events were reconstructed using an interactive graphics system.  All recognizable tracks and showers in the event were individually measured yielding momenta and energies.  Protons were identified by their high ionization and lack of Coulomb scattering.  Finally the individual particle momenta and energies were added to form the incoming neutrino four-vector.

     The topology classes provide a crude energy separation.  The track and shower samples have a lower average neutrino energy than the multiprong sample and both are lower than the mean PCE energy.  Histograms of the Monte Carlo generated neutrino energy, $E_\nu$, for the three classes are shown in Fig.~\ref{fig:energy} for the \muflvr and \eflvr events.  Note that the contained single shower distribution extends to higher energies than the track sample because of the better containment of high energy showers.

     The numbers of events analyzed are shown in Table~\ref{eventno}.  The oscillation analysis described in this paper imposed a minimum 300 MeV/$c$ cut on the lepton momentum for LoRes track and shower events and the total visible momentum for LoRes multiprong events.  The numbers headed ``Raw Events'' in the table are those for the total event sample without this cut, the other numbers include the cut. The table shows that the rock background is concentrated in low energy events which are removed by the cut.  The value of the cut was chosen to optimize the analysis sensitivity by reducing the background component while retaining the neutrino signal. The final two columns are the fitted number of qs-rock events and the number of neutrino events after subtraction of the qs-rock background as described in Sect.~\ref{sec:backsub}.

     More details of the data sample and the analysis procedures are given in Ref.~\cite{mayly}.

\begin{table}[hbt] 
\centering 
 
{\begin{tabular}{|c|c||c|c|c||c|c|c|c|c|} 
\hline 
Event & Fla- &\multicolumn{3}{c||}{ Raw  Events} & \multicolumn{5}{c|}{ 300 MeV/$c$ cut} \\ 
Class & vor  & qs-data & MC & rock & qs-data&  MC &rock &qs-rock & neutrino\\ 
\hline 
FCE HiRes&$\mu$ & 114 &1149& 73 &114 & 1115.1 & 73 & 12.1$\pm$6.9 & 101.9$\pm$12.7 \\ 
FCE HiRes& e & 152 &1070& 69 &152 & 1047.4 & 69 & 5.3$\pm$2.1 & 146.7$\pm$12.5 \\ 
FCE LoRes& $\mu$ & 148 &900 &406 & 61 &  457.5 & 77 & 11.5$\pm$6.2 & 49.5$\pm$9.9 \\
FCE LoRes& e  & 177 &850 &704 & 71 &  402.5 &176 &14.0$\pm$4.6 &  57.0$\pm$9.6 \\ 
PCE    &$\mu$   &  53 &373 & 11 & 53 &  384.3 & 11 & 0.3$\pm$0.9 &52.7$\pm$7.3 \\
PCE    & e    &   5 & 51 &  0 &  5 &   51.5 &  0 & 0.0$\pm$0.1 & 5.0$\pm$2.2 \\ 
NC+ambig & & 46 &246 &190 & 32 &  165.7 &110 & 7.6$\pm$6.7 & 24.4$\pm$8.8 \\ 
\hline 
Total  &   &     695 &4639&1453& 488 & 3624.0 &516 & 50.8 & 437.2 \\ 
\hline 
\end{tabular}} 
\caption{Event samples in the 5.90 fiducial kiloton-year exposure. The columns for raw events are the total numbers of events reconstructed in this experiment. The qs-data, MC and rock classes are defined in Sects.~\ref{sec:evclass} and \ref{sec:mclass}. The columns headed `` 300 MeV/$c$ cut '' give the event numbers used in the oscillation  analysis described in this paper. The MC numbers in these columns have been weighted to convert from the Bartol 89 flux prediction to the Bartol 96 values. The qs-rock column gives the fitted number of background qs-rock events and the neutrino column is the number of neutrino qs-data events after subtraction of the qs-rock background as described in Sect.~\ref{sec:backsub}.
}
\label{eventno} 
\end{table}

\section{Evidence for \numu flavor disappearance}
\label{dataa}

     The neutrino oscillation parameters were determined using an unbinned maximum likelihood analysis of the complete data set based upon the Feldman-Cousins procedure, as described in Sect.~\ref{oscanal}.  However it is instructive first  to examine subsets of the data to observe the effects of oscillations.

\subsection{Flavor Ratio-of-Ratios}
\label{ratio} 

\begin{figure}[htb]
\centerline{\epsfig{file={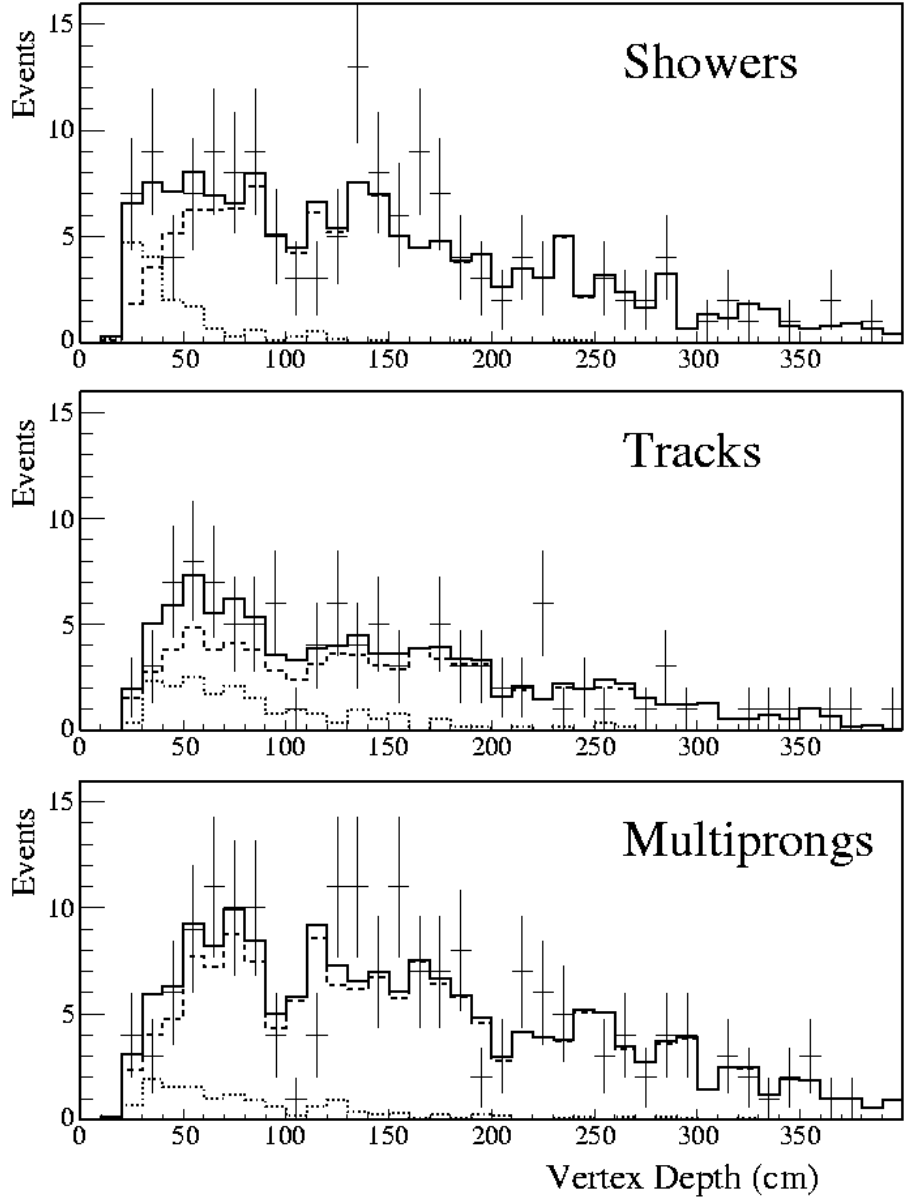} ,width=12.0cm}} 
\caption{Depth distributions for shower (top), track (center), and multiprong events (bottom) after the 300 MeV/$c$ cut.  The points with error bars are the qs-data events. The dashed histograms are the unoscillated Monte Carlo events. The dotted histograms are the rock events and the solid histograms are the sum of Monte Carlo and rock.  The fraction of qs-rock is determined from a fit of the Monte Carlo plus rock events to the qs-data event distribution. The summed distribution is normalized to the qs-data events.
}
\label{fig:depth}
\end{figure}

  The first indication of neutrino oscillations in atmospheric neutrinos came from the measurement of the ratio-of-ratios, $R_\nu$ \cite{Kam}, defined here as

\begin{eqnarray}
R_{\nu}=\frac{(\nu_{\mu}/\nu_{e})_{data}}{(\nu_{\mu}/\nu_{e})_{MC}}.
\label{eq:rnu}
\end{eqnarray}

Two corrections, for qs-rock background and for flavor misidentification, are required to the raw numbers of qs-data \muflvr and \eflvr events.
\subsubsection{Correction for qs-rock background}
\label{sec:backsub}  
For the measurement of $R_\nu$, the number of qs-rock background events contained in the qs-data sample was estimated by  fitting the depth distribution of the qs-data events to a sum of the MC and rock depth distributions. The extended maximum likelihood method of Ref.~\cite{barlow} was used, which allows for the finite statistics of the Monte Carlo and rock distributions. The depth was defined as the distance from the event vertex to the closest surface of the detector, excluding the floor.   Neutrino events are expected to be distributed uniformly throughout the detector while the neutron and gamma induced events are attenuated by their respective interaction lengths.  The depth distributions are shown in Fig.~\ref{fig:depth}.  The fits are made to the  track,  shower and multiprong samples separately since the backgrounds are significantly different in each sample.  The excess of events at low depths is clear in the shower and track samples.  The multiprong sample has little qs-rock background. The background in the track and multiprong samples contains only neutron induced events and the rock events  are attenuated according to the 80 cm neutron attenuation length. The shower background  has an additional component due to gammas with a 15 cm attenuation length.  Table~\ref{eventno} shows the number of neutrino events in the 300 MeV/$c$ cut sample after background subtraction.  It can be seen that the background in the high energy PCE sample is very small.  

In this section the fitted amount of qs-rock background is not correlated with the oscillation parameters, \sqdmns, as it is determined only from the depth distribution. In the full oscillation analysis, described in Sect.~\ref{oscanal}, the fit includes the additional information of the MC and rock \loe distributions and the relative \muflvr to \eflvr normalization.

\subsubsection{ Correction for event misidentification}

  Table~\ref{tab:flavor_matrix} shows the identification matrix determined from the MC truth and assigned flavor for fully contained events only, after the 300 MeV/$c$ cut. The wrong flavor contamination of the \muflvr and \eflvr events is 3.8\% and 2.8\% respectively.  There is also a contamination of neutral current events of 7.4\% and 6.8\% respectively. The misidentification of partially contained events is negligible.

\begin{table}[hbt] 
\centering 
{\begin{tabular}{|l|ccc|}
\hline
  MC        &      &Assigned flavor& \\
 Truth      & $\mu$&   e  & NC \\ 
\hline 
  $\nu_\mu$ & 1396.8 & 41.0   & 50.8 \\ 
  $\nu_e$   &   59.0 & 1311.9 & 40.5 \\ 
 NC         &  116.8 & 98.0 & 45.5 \\ 
\hline 
\end{tabular}}
\caption{Flavor identification matrix after the 300 MeV/$c$ cut from ``blind" processing
of fully contained $\nu$ MC events interspersed throughout the data. The NC column includes events classified as having ambiguous flavor as well as those definitively classified as NC. The events are weighted to correspond to the Bartol 96 flux.}
\label{tab:flavor_matrix}
\end{table} 

\subsubsection{Ratio-of-ratios results}

For comparison with previous experiments the ratio-of-ratios is first quoted for fully contained events.  The maximum sensitivity is obtained using the 300 MeV/$c$ cut sample to reduce the effects of the background subtraction.

The raw data ratio-of-ratios, $R$, is defined as
\begin{equation}
R=\frac{(D_{\mu}/D_{e})}{(C_{\mu}/C_{e})}
\end{equation}

\noindent where $D_{\mu}$ and $D_{e}$ are the numbers of background subtracted qs-data \muflvr and \eflvr events respectively, listed in the last column of Table~\ref{eventno}, and  $C_{\mu}$ and $C_{e}$ are the numbers of \muflvr and \eflvr MC events.
The result is

\begin{eqnarray}
R=0.69 \pm 0.10(stat) \pm 0.06(syst).
\end{eqnarray}
\noindent The systematic errors are discussed in Ref.~\cite{Allison2}.

 The fraction of $\nu_{\mu}$ remaining after oscillations, $R_{\mu}$, and $A$, the 
normalization of the experiment relative to the Bartol 96 flux, can be determined  using  the identification matrix in Table~\ref{tab:flavor_matrix}. On the assumption that only \numu oscillate, $R_{\mu}$ is equivalent to  $R_\nu$, the corrected ratio-of-ratios defined in Eq.~(\ref{eq:rnu}).

The numbers of \muflvr and \eflvr events are given by
\begin{eqnarray}
D_{\mu}=A(R_{\mu}T_{\mu}+T_{e}+T_{n})C_{\mu} \\
D_{e}=A(R_{\mu}S_{\mu}+S_{e}+S_{n})C_{e}
\end{eqnarray}

\noindent where $T_{\mu}$, $T_{e}$, $T_{n}$ are the probabilities for 
\muflvr events to be $\nu_{\mu}$, $\nu_{e}$, or NC as obtained from the 
identification matrix  and $S_{\mu}$, $S_{e}$, $S_{n}$ are those probabilities for 
\eflvr events.

Dividing the two equations and noting that $T_{e}+T_{n}=1-T_{\mu}$ and 
$S_{e}+S_{n}=1-S_{\mu}$, a relation between $R_{\mu}$ and $R$ is obtained:
\begin{eqnarray}
R_{\mu}=\frac{(1-S_{\mu})R-1+T_{\mu}}{T_{\mu}-S_{\mu}R}
\end{eqnarray}

\noindent yielding
\begin{eqnarray}
 R_{\mu} & = & 0.64 \pm 0.11 \pm 0.06 \\
         & = & R_{\nu} \nonumber
\end{eqnarray}
\noindent and also giving a value of $A$ corresponding to 86\% of the Bartol 96 flux.

A trend in the variation of $R_\nu$ with energy can be seen by comparing $R_\nu^{ts}$ for the track and shower samples alone, which have a lower average energy and $R_\nu^{all}$ for the full sample including the PCE, which has a higher average energy, as shown in Fig.~\ref{fig:energy}.  The values are
\begin{eqnarray} 
R_\nu^{ts}=0.51 \pm 0.13
\end{eqnarray} 
\noindent with $A$ at 89\% of Bartol 96 and
\begin{eqnarray}
R_\nu^{all}=0.72 \pm 0.10
\end{eqnarray}
\noindent with $A$ at 85\% of Bartol 96.

The value of $R_\nu$ increases with 
increasing average energy of the sample, as expected if the 
deviation from 1.0 is due to oscillations. The most significant 
deviation is for the lower energy single track/shower sample, 
an effect of more than three standard deviations.

The value of $A$ for the full data set, $85 \pm 7$ \% of the Bartol 96
prediction, may be compared to the value obtained in the more detailed likelihood analysis which includes the \loe information and is described in Sect.~\ref{oscanal}.

\begin{figure}[htb]
\centerline{\epsfig{file={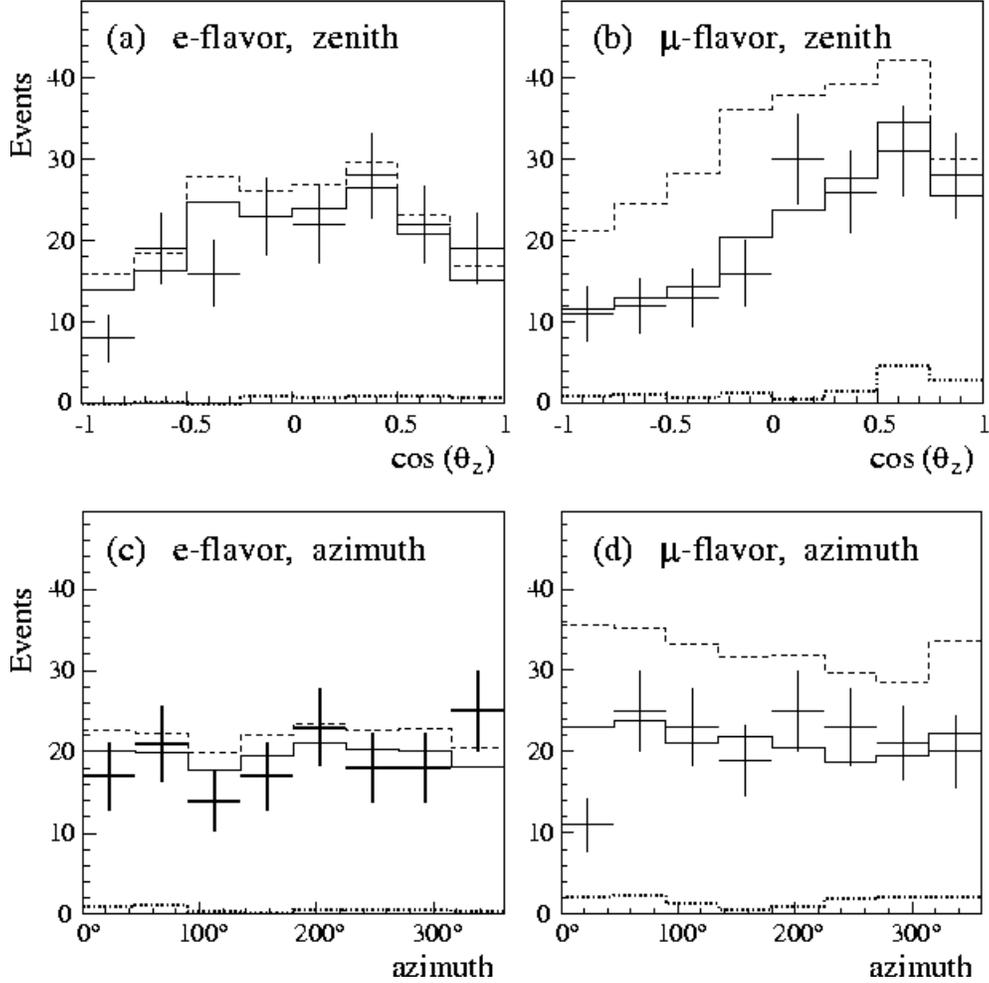} ,width=13.0cm}} 
\caption{Angular distributions for HiRes \eflvr events (plots (a) and (c)) and \muflvr events  (plots (b) and (d)).  Plots (a) and (b) show the cosine of the zenith angle and plots (c) and (d) the azimuth angle.  The points with error bars are the qs-data. The dashed histograms are the sum of the predicted unoscillated neutrino distribution plus the fitted qs-rock contribution. The solid histograms are the same but with the neutrino distribution weighted by the oscillation probability predicted by the best fit parameters from the analysis described in Sect.~\ref{oscanal}. The dotted histograms are the contribution of the qs-rock background. Downward going events have cos$\theta_z$=+1.0. Note the depletion of \muflvr events at all but the highest value of cos$\theta_z$.
}
\label{fig:azizen}
\end{figure}

\subsection{Angular and L/E distributions}
   In this analysis pure two flavor \numutonutau oscillations will be assumed, based on the absence of observed effects in the \eflvr events, shown below, and the results of the CHOOZ \cite{CHOOZ} and Super-K \cite{Super-K_OSC} experiments.  Two flavor oscillations in vacuum leading to \numu disappearance are described by the well-known formula:

\begin{eqnarray}
P(\nu_{\mu} \rightarrow \nu_{\mu}) = 1.0 -
\sin^2 2\theta \cdot \sin^2 \left[1.27 \mbox{ } \Delta m^{2}
[\mbox{eV}^2] \cdot \left(\frac{L \mbox{[km] }}{E \mbox{[GeV] }}\right) \right].
\label{eq:prob}
\end{eqnarray}

\noindent The \numu survival probability, $P$, is a function of \loe where $L$ is the distance traveled by the neutrino and $E$ is the neutrino energy.  The parameters to be determined are \delmns, the mass squared difference, and \sqsinns, the maximum oscillation probability.

The distance, $L$, is calculated by projecting the measured neutrino direction back to the atmosphere. It is related to the zenith angle, $\theta_z$, by

\begin{eqnarray} 
L(\theta_z) = 
\sqrt{(R_e - d_d)^2 \cos^2\theta_z + (d_d + h)(2R_e - d_d + h)} 
- (R_e - d_d)\cos\theta_z 
\end{eqnarray}

\noindent where $R_e$ is the radius of the Earth, $d_d$ is the depth of the detector and $h$ is the neutrino production height in the atmosphere. 
 The mean of the production height distribution for neutrinos of a given  type, energy and zenith angle \cite{Ruddick} was used as the estimator of $h$. The variation in production height is comparable to the path length for neutrinos coming from above the detector. 
The distance travelled is a function of $cos\theta_z$ but not of the azimuth angle. Oscillation effects are therefore expected in the shape of the zenith angle distribution but not in that of the azimuth angle distribution.

Fig.~\ref{fig:azizen} shows the azimuth and zenith angle dependence of the HiRes  \muflvr and \eflvr events.  The effects of oscillations are most visible in the HiRes sample.  The points with errors are the qs-data. The  dashed histograms are the expected, unoscillated, MC neutrino contribution based on the Bartol 96 flux calculation plus the fitted qs-rock contribution. The solid histograms are the same but with the MC  weighted by the oscillation probability predicted by the best fit oscillation parameters of the analysis described in Sect.~\ref{oscanal}. The dotted  histograms are the fitted background qs-rock contribution. It can be seen that both the azimuth and zenith angular distributions for the \eflvr events (Figs.~\ref{fig:azizen}a and \ref{fig:azizen}c) are consistent with the unoscillated MC prediction up to a 10-15\% normalization of the overall flux.  On the other hand, the \muflvr zenith angle distribution (Fig.~\ref{fig:azizen}b) shows a 50\% deficit at large zenith angles (large $L$) but little deficit for downward going events (small $L$), confirming the observation of similar effects in the Super-K experiment.  Note that at the high magnetic latitude of Soudan the azimuth angle distribution of both flavors (Figs.~\ref{fig:azizen}c and \ref{fig:azizen}d) is predicted and observed to be flat, unlike the distribution at Kamioka where there is a pronounced East-West asymmetry.

\begin{figure}[htb]
\centerline{\epsfig{file={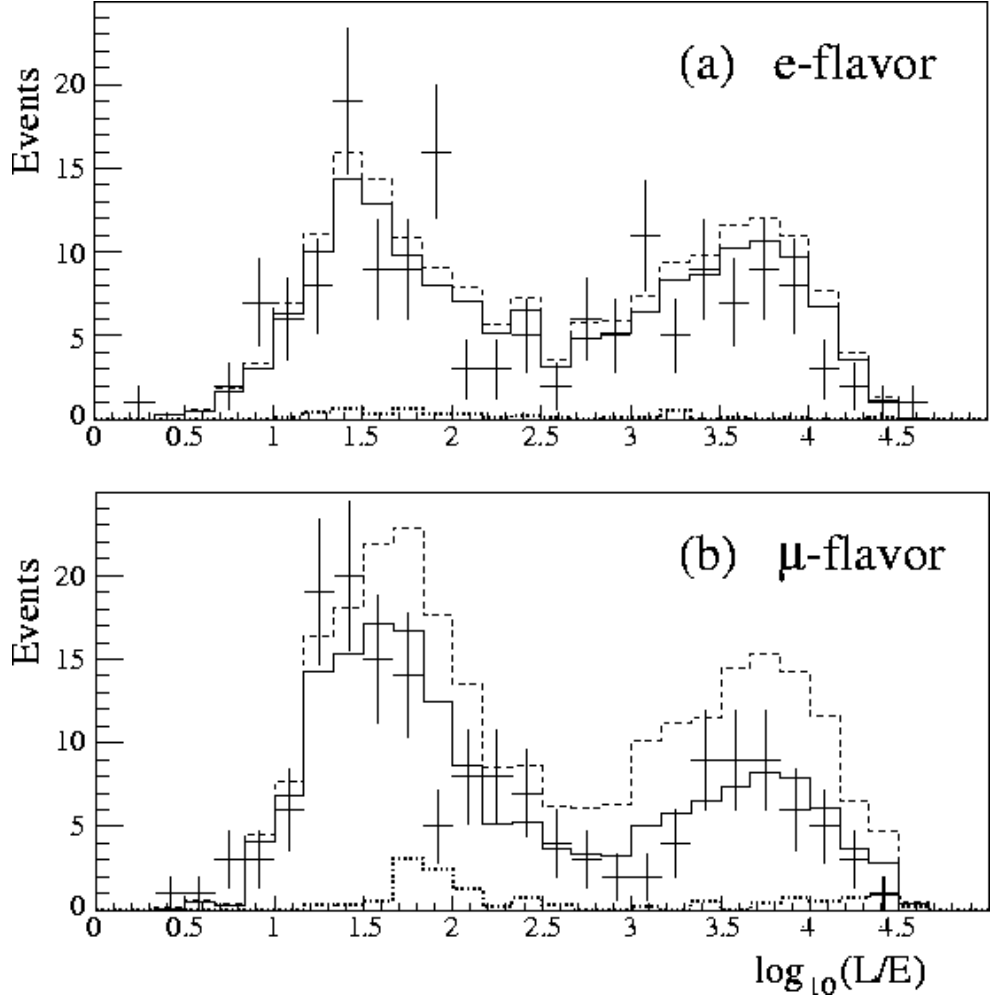} ,width=13.0cm}} 
\caption{The HiRes \ltenloe distribution for \eflvr events (top) and \muflvr events  (bottom).  The points with errors are the qs-data. The dashed histograms are the prediction of the unoscillated Monte Carlo plus the fitted qs-rock contribution.  The solid  histograms are the same but with the Monte Carlo weighted by the best fit oscillation parameters from the analysis described in Sect.~\ref{oscanal}.  The dotted histograms are the contribution of the qs-rock background.  A depletion of \muflvr events above values of \ltenloe of approximately 1.5 can be seen.
}
\label{fig:lovere}
\end{figure}

Fig.~\ref{fig:lovere} shows the \ltenloe distributions for the HiRes \eflvr and \muflvr samples. The double peaked structure is a geometrical effect, reflecting the spherical shape of the Earth. The peak at lower \ltenloe consists  predominantly of downward-going neutrinos from the atmosphere above the detector, while the peak at higher \ltenloe contains upward-going neutrinos from the other side of the Earth.  Again the \eflvr sample follows well the MC prediction, up to a normalization factor, indicating that within the errors of this experiment  there is no evidence for \nue oscillations.   The \muflvr  sample shows a deficit of events above a \ltenloe value of around 1.5.  Below this value there is little, if any, loss of events.  This implies an upper limit on the value of \delm of about 0.025 eV$^2$ which is reproduced in the detailed fits described in Sect.~\ref{oscanal}.

\section{Neutrino Oscillation Analysis}
\label{oscanal}

An extended maximum likelihood  analysis assuming two-flavor $\nu_\mu \rightarrow \nu_\tau$ 
oscillations has been used to obtain estimates of the neutrino oscillation parameters. The  significance of the result and the
confidence intervals on the oscillation parameters are determined using the unified
method advocated by Feldman and Cousins~\cite{Feldman_Cousins}. 
\subsection{The likelihood function}
\label{sec:eml}
The likelihood function used to describe the qs-data assumes that the  sample  is composed of neutrino interactions, represented by the Monte Carlo events, and qs-rock background events, represented by the rock sample.  Each sample is divided into \muflvrns, \eflvr and NC plus ambiguous events. Since neither \nue or NC events are assumed to oscillate they can be considered as a single category.  For shorthand in the following they are combined under the heading of \eflvr. 

The \loe distribution of the \muflvr events is examined for evidence of oscillations.   The total number of  events, \muflvr plus \eflvrns, provides the normalization of the Monte Carlo exposure. Monte Carlo events with misidentified flavor are included in the \muflvr or \eflvr samples with the oscillation probabilities appropriate to their true parameters. Charged current interactions of  $\nu_\tau$ were not generated in the Monte Carlo.  At the Super-K best fit oscillation parameters, approximately two interactions producing $\tau$ leptons are expected in this data sample. 

Each event in the \muflvr sample is characterized by its measured values, 
$(x_i,d_i)$, of $x \equiv$ \ltenloe and depth within the detector, $d$. The true value of 
$L/E \equiv (L/E)^{\rm true}$ is also known for each  event in the Monte Carlo 
sample.
The distinction between the five categories of \muflvr event (HiRes tracks and multiprongs, LoRes tracks and multiprongs and PCEs) is maintained for the analysis.

The log-likelihood function used is:

\begin{eqnarray}
{\cal{L}} & = & \sum_{k=1,5} \left\{ \sum_{i=1,N_\mu^k} \ln Q^k(x_i,d_i)
+  N_\mu^k \ln M_\mu^k - M_\mu^k \right \}+ N_e \ln M_e - M_e.
\label{eq:logl}
\end{eqnarray}

\noindent The function $Q^k(x_i,d_i)$ is the normalized joint $(x_i,d_i)$ probability density function (pdf) for category $k$ and the $k$-summation is taken over the five \muflvr event categories. The symbols
$N_\mu^k$ and $N_e$ denote the total number of qs-data events in the \muflvr and \eflvr categories and $M_\mu^k$ and $M_e$ are the predicted number of events, {\it i.e.} the sum of MC neutrino plus qs-rock events. The function  $Q$ represents the shape information in the \loe and depth distributions and the other terms arise from the data normalization.
 
The joint $(x,d)$ distribution
of the qs-data is the sum of the joint $(x,d)$ distributions of true neutrino 
events and qs-rock background events. The simplification is made that the joint distributions can be represented
as the product of the  distributions of $L/E$ and depth, {\it i.e.\/} that 
there is no correlation between the $L/E$ of an event and its depth in the detector.
Thus the $(x,d)$ pdfs for \muflvr neutrino and qs-rock events are 
$X^k_\mu(x_i)D^k_\mu(d_i)$ and $X_R^k(x_i)D_R^k(d_i)$ respectively. 
 $X^k_\mu(x_i)D^k_\mu(d_i)$ is the normalized probability that event 
$i$ of \muflvr category $k$
has a value of \ltenloe $= x_i$ and is found at a depth $d_i$ if it is a neutrino 
interaction, and $X_R^k(x_i)D_R^k(d_i)$ is the corresponding probability
that the event arises from a rock background interaction.  Substituting for $Q$ and normalizing it to 1.0, the likelihood function becomes

\begin{eqnarray}
{\cal{L}} & = & \sum_{k=1,5} \left\{
\sum_{i=1,N_\mu^k} \ln \left[{{AC^{k} X^k_\mu(x_i)D^k_\mu(d_i)
+N_R^k X_R^k(x_i)D_R^k(d_i)}\over{M_\mu^k}} \right]
+N_\mu^k \ln M_\mu^k - M_\mu^k
\right\}\nonumber\\
 & & + N_e \ln M_e - M_e \,.
\label{eq:klogl2}
\end{eqnarray}
  The factor $A$ is a free parameter representing the normalization of the MC sample to the qs-data and  $C^{k}$ is the oscillated
number of \muflvr $\nu$ MC interactions for a given pair of oscillation 
parameters (\sqdmns). $N_R^k$ is the  number of qs-rock background events in 
\muflvr category $k$. The total expected number of \muflvr qs-data  events in category $k$ is
\begin{eqnarray}
 M_\mu^k = AC^{k} + N_R^k.
\end{eqnarray}
\noindent where
\begin{eqnarray}
C^k & = &  
\sum_{j = 1,C^{k}_0} P((L/E)_{j}^{\rm true})
\end{eqnarray}
\noindent and  $C^{k}_0$ is the total unoscillated number of MC events of category $k$. The oscillation survival probabilities, $P((L/E)^{\rm true})$ as defined in Eq.~(\ref{eq:prob}), are computed using the known, true, 
values of $L/E$ and the given oscillation parameters.   The expected number of qs-data \eflvr events is given by
\begin{eqnarray}
M_e=AC^e+N_R^e
\end{eqnarray}
\noindent where $C^e$ is the total number of Monte Carlo \eflvr events and $N_R^e$ is the number of \eflvr qs-rock events.

\subsection{Probability density functions}
\label{sec:pdf}

Continuous pdfs, $X^k_\mu(x)$ and $X_R^k(x)$, 
 are constructed from the Monte Carlo and rock
samples
to represent the expected neutrino and qs-rock background  distributions for each category of \muflvr events. 
The use of a continuous pdf is preferred over the more conventional histogram
representation since it does not require any arbitrary choice of binning. 
The 
pdfs are constructed by spreading the positions of the 
measured $x$ values of the Monte Carlo and rock events with smooth functions. This enables a continuous functional form of the pdf to be obtained 
from the finite set of discrete parameter values of the MC events.
Explicitly the pdfs are constructed as follows:
\begin{eqnarray}
X^k_\mu(x) & = &  {1 \over C^k} 
\sum_{j = 1,C^{k}_0} g^k_\mu(x_{j} - x) P((L/E)_{j}^{\rm true})
\label{eq:xpdf} 
\end{eqnarray}
where $g^k_\mu(x_{j} - x)$ is the spreading function. Likewise
\begin{eqnarray}
X^k_R(x) & = &  {1 \over N^{k}_{R0}} 
\sum_{j = 1,N^{k}_{R0}} g^k_R(x_{j} - x) 
\end{eqnarray}
where $N^{k}_{R0}$ is the total number of events in the category $k$ rock sample.
 A normalized
Gaussian form is chosen for the spreading functions,
\begin{eqnarray}
g^k(x_j - x) & = & 
{1 \over{{\sqrt{2\pi} \sigma_k}}} \exp( -( x_j - x)^2/2 \sigma_k^2)\,.
\label{eq:gkern}
\end{eqnarray}
The pdfs constructed according to Equations~\ref{eq:xpdf}--\ref{eq:gkern} are
normalized to unity. A different value of $\sigma_k$ is used for each different
event category. The value of $\sigma_k$ is chosen to provide a
representation of the pdf without statistical dips which could be mistaken for oscillation structures.  The value  is a balance between small values which emphasize the resolution of the experiment and larger values which smooth  the finite statistics of the samples. The values
of $\sigma_k$ used are given in Table~\ref{tab:sig}.

The depth pdfs $D^k_\mu(d)$ and $D_R^k(d) $ are represented by  histograms of the depth distributions  similar to those shown in Fig.~\ref{fig:depth}.
 
\begin{table}[hbt] 
\centering 
{\begin{tabular}{|l|c|c|}
\hline
         Event category & $\sigma$ MC     & $\sigma$ rock \\
\hline 
 HiRes tracks & 0.075 & 0.12 \\
 LoRes tracks & 0.110 & 0.13 \\
 PCE          & 0.100 & 0.25 \\
 LoRes multiprongs & 0.180 & 0.25 \\
 HiRes multiprongs & 0.100 & 0.25 \\
\hline 
\end{tabular}}
\caption{Values of $\sigma$ used to construct the pdfs.}
\label{tab:sig}
\end{table}

\subsection{Determination of the oscillation parameters}
\label{sec:bestfit}

  The two oscillation parameters to be determined are \sqdm for which proper coverage is provided. Seven other unknown quantities, the normalization of the MC, $A$, the five rock fractions for the different $\mu$ flavor samples and the rock fraction for the \eflvr sample are nuisance parameters.  In general their fitted values will be correlated with the values of the oscillation parameters.  The qs-rock background in the PCE sample is known to be very small from the small number of partially contained rock events in Table~\ref{eventno}, the measured shield efficiency and the measured neutron energy distribution.  It is set to zero in this analysis. The amount of qs-rock background in the \eflvr sample is estimated by a fit to the \eflvr depth distributions as described in Sect.~\ref{sec:backsub} and is assumed to be independent of the oscillation parameters.

The negative log likelihood is calculated on a 15 $\times$ 80 grid of \sqsin $\times$ log$_{10}(\Delta m^2)$ with \sqsin between 0.0 and 1.0 and \delm between $10^{-5}$ and $10^0$ eV$^2$.   The \numu survival probability, $P$, defined in Eq.~(\ref{eq:prob}), is averaged over the area of each grid square. The range of \delm was chosen such that outside this range the predictions for \ltenloe are constant to a good approximation.  Below the lower limit the survival probability is close to one for the whole \ltenloe range, above the upper limit the probability averages to 0.5.  

At each grid square the likelihood is minimized as a function of each of the remaining four \muflvr rock fractions.  The value of $A$ is calculated by requiring that the predicted number of Monte Carlo plus qs-rock events equals the number of qs-data events.  For simple likelihood functions this is a mathematical condition for the minimum. It was tested and found to be a very good approximation for this likelihood function.

\begin{figure}[htb]
\centerline{\epsfig{file={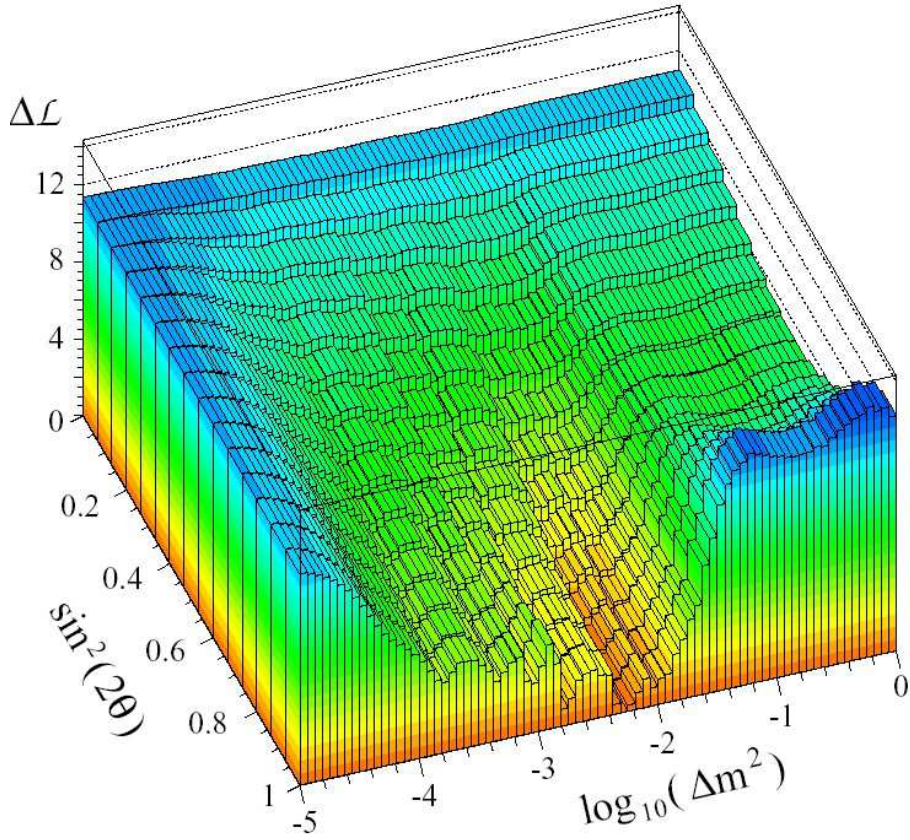} ,width=12.0cm}} 
\caption{The data likelihood difference, $\Delta \cal{L}$, plotted as a function of \sqsin and log$_{10}(\Delta m^2)$.}
\label{fig:datallh}
\end{figure}

The lowest negative log likelihood on the grid is found and the difference between that and the value in each \sqdm square ($\Delta \cal{L}$) is plotted in Fig.~\ref{fig:datallh}.  The resulting surface
 exhibits a broad valley which curves from a mean value of \sqsin of around $0.5$ at high \delm to \delm between $10^{-4}$ and $10^{-2}$ eV$^2$ at high \sqsinns.  This is the locus of constant $R_\nu$ as defined in Eq.~(\ref{eq:rnu}).  The shape information in the \loe distribution favors the high \sqsin region.  The best likelihood occurs for the grid square centered at \delm = 0.0052 eV$^2$, 
        \sqsin~=~0.97. This grid square will be referred to as the best fit point.  The value of $A$ is 90\% of the Bartol 96 prediction and the total number of \muflvr qs-rock events is 16.8, 9.6\% of the fully contained \muflvr sample.  The flux normalization is discussed in more detail in Sect.~\ref{flux}.

Since the likelihood at each grid square is an average over the area of the square, there is no \sqsin = 0 point in the analysis.  The grid square with the lowest values of \sqsin and \delm  is taken as a good approximation to the case of no oscillations.  The likelihood rise from the minimum at this grid square is 11.3. The hypothesis of no oscillations is thus strongly disfavored.  A quantitative assessment of its probability is presented in Sect.~\ref{sec:likprob}.

 Fig.~\ref{fig:datallh} shows that the likelihood surface is not parabolic in the region of the minimum.  Errors on the parameters cannot be accurately defined using a simple likelihood rise.  Confidence level contours have  been determined using the method of Feldman and Cousins as described in Sect.~\ref{FC}.

\begin{figure}[htb]
\centerline{\epsfig{{file=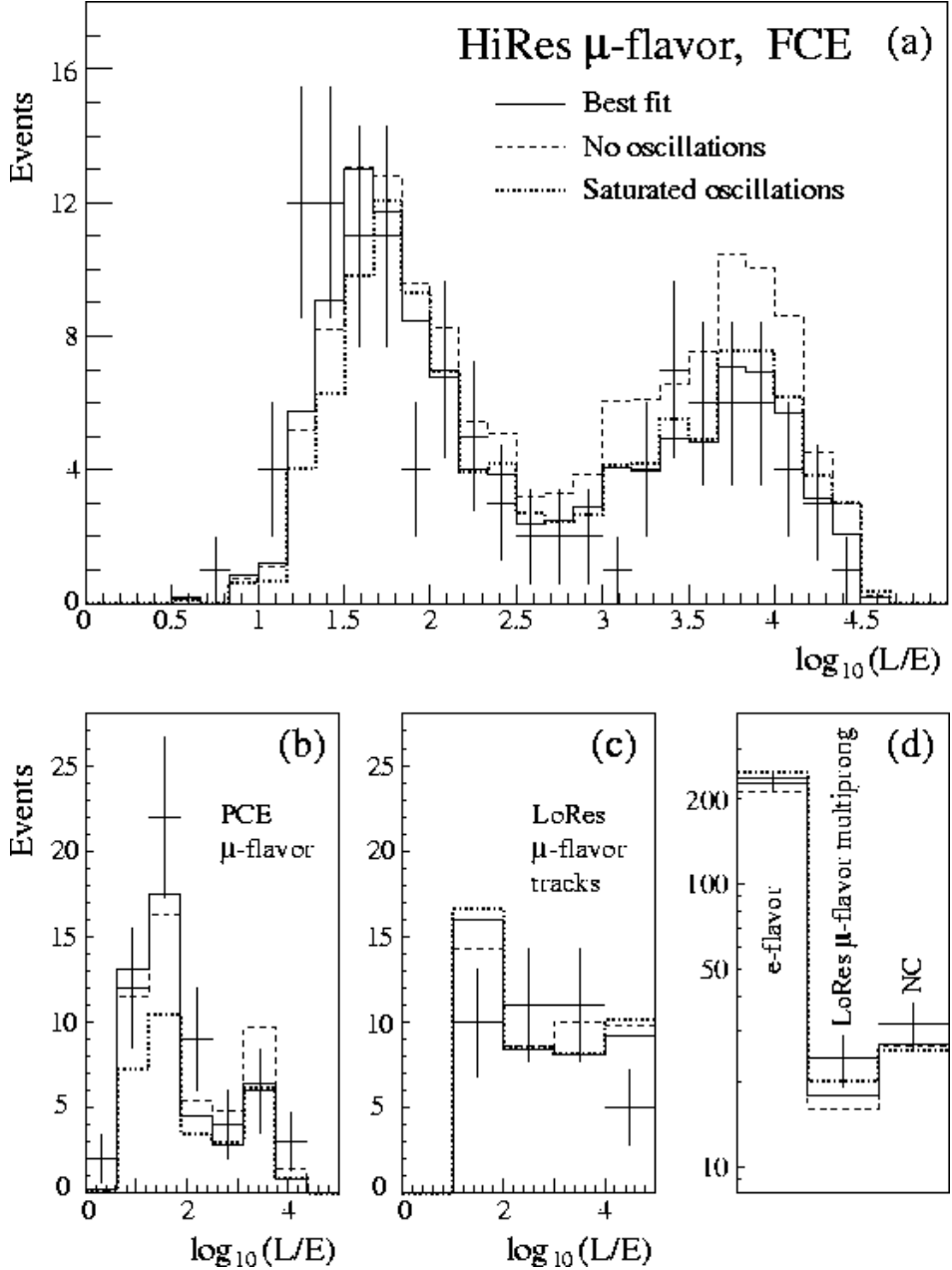},width=12.5cm}}  
\caption{
Comparison of the qs-data with the fit predictions.  Each plot shows the qs-data as points with error bars. The sum of the neutrino Monte Carlo plus qs-rock for the best fit point in \sqdm is the solid histogram. The case of no oscillations is the dashed histogram and saturated oscillations the dotted histogram.  Plot (a) is the \ltenloe distribution for the HiRes \muflvr FCE.  Plot (b) is the \ltenloe distribution for the \muflvr PCE. Plot (c) is the \ltenloe distribution for the  LoRes \muflvr tracks. Plot (d) shows on a log scale the total events for the \eflvr sample (not including the NC plus ambiguous events), the LoRes \muflvr multiprongs and the NC plus ambiguous flavor events.  The HiRes \muflvr FCE and PCE events which are summed in Fig.~\ref{fig:lovere} are shown separately here.
}
\label{fig:data}
\end{figure}

\subsection{Comparison with binned data}

The results of an unbinned  likelihood analysis are difficult to visually compare with the data.  To provide such a comparison, 
Fig.~\ref{fig:data} shows the qs-data plotted together with  the histogram of the best fit prediction of the sum of oscillated neutrino Monte Carlo plus qs-rock background. Also histogrammed are the predictions for no oscillations  (dashed histograms) and for \sqsinns=1.0, \delmns=1.0 eV$^2$ (dotted histograms), referred to hereafter as ``saturated oscillations''. Although the oscillations in \ltenloe are rapid at this point, averaging to close to 0.5, the saturated oscillation histogram is not just half the no oscillation histogram because of the differences in the fitted normalization and amounts of rock background.  The bin sizes in \ltenloe are chosen to be appropriate for the statistics, resolution and sensitivity to oscillations.  Fig.~\ref{fig:data}a shows the HiRes FCE data.  Note that the no oscillation hypothesis fits poorly to the high \ltenloe peak and saturated oscillations do not represent the low \ltenloe values.  Fig.~\ref{fig:data}b shows the PCE data. Again the saturated oscillation hypothesis gives a bad fit to the low \ltenloe points. Fig.~\ref{fig:data}c shows the LoRes track events.  The resolution in \ltenloe is poor for this sample and there is not much discrimination between oscillation hypotheses. Fig.~\ref{fig:data}d shows the summed number of events in the remaining three categories (\eflvrns, LoRes \muflvr multiprongs and NC plus ambiguous events) where there is no detectable sensitivity to oscillations in their \ltenloe distribution.  Note that the normalization of the neutrino Monte Carlo is strongly constrained by the number of \eflvr events.  The variation in the predicted number of \eflvr events for the three hypotheses represents the difference in normalization of the flux required to give the best fit to these hypotheses. In the saturated oscillation case more qs-rock events are added to compensate for the neutrino events lost by oscillations.

Table~\ref{tab:chisq} shows a $\chi^2$ comparison for the best fit, the fit with the current Super-K best fit parameters (\delm = 0.0025, \sqsin = 1.0) \cite{SuperKallowed}, the fit with no oscillations and the fit with saturated oscillations.  Bins are combined to give a minimum of five events per bin for the $\chi^2$ calculation.   

\begin{table}[hbt] 
\begin{tabular}{|l|c|c|c|c|c|c|}
\hline
  Case       & HiRes FCE $\mu$  & LoRes FCE $\mu$ & PCE $\mu$ & \eflvr & NC & Total \\ 
\hline 
 Best fit              & 14.1 & 10.1 & 3.8 & 0.2 & 0.8 & 29.0 \\
 No oscillations       & 42.9 & 9.7 & 3.9 & 1.4 & 0.9 & 58.9 \\
 Saturated oscillations  & 22.8 & 11.5& 13.3 & 1.4 & 1.3& 50.3 \\
 Super-K best fit      & 15.6 & 10.3 & 3.3 & 0.1 & 0.7 & 30.0 \\
\hline 
 Number of bins        & 14   & 5 & 4 & 1 &1 &25 \\
\hline
\end{tabular}
\caption{The $\chi^2$ for the comparison of various \sqdm predictions to binned qs-data. The four \muflvr background fractions and the normalization parameter, $A$, are variables in each case.} 
\label{tab:chisq}
\end{table}

The $\chi^2$ comparison is in good agreement with what can be deduced from the likelihood surface.
The best unbinned likelihood fit parameters give a good $\chi^2$ comparison of the binned qs-data to the prediction.  The Super-K best fit parameters also give a good, though not the best, fit to the qs-data. The no and saturated oscillation hypotheses are strongly disfavored.  However the systematic errors and non-Gaussian effects included in the Feldman-Cousins analysis are not included in this $\chi^2$ comparison.  A full analysis of the no oscillation hypothesis is given in Sect.~\ref{sec:likprob}.

\section{Determination of the Confidence Regions}
\label{FC}
\subsection{Feldman-Cousins analysis}

   If all errors were Gaussian, if there were no systematic effects and if there were no physical boundaries on the parameters, a 90\% confidence contour in \sqdm could be obtained from the data likelihood plot shown in Fig.~\ref{fig:datallh} by taking those grid squares where the likelihood rose by 2.3 above the minimum value.  However this is far from the case in this analysis.  The values of \sqsin are bounded by 0.0 and 1.0 and the best fit is close to the upper bound.  The errors on \loe are a complicated function of the measurement errors and there are systematic errors to be taken into account.

The procedure proposed by Feldman and Cousins is a frequentist approach which uses a Monte Carlo method of allowing for these effects \cite{Feldman_Cousins}.  In their method MC experiments are generated and analyzed  at each grid square on the \sqdm plane.  These experiments have the statistical fluctuations appropriate to the data exposure and can have systematic effects incorporated.

In this analysis each MC experiment was generated by selecting a random sample of the MC and rock events from the total sample of these events.  The normalization of the MC neutrino events was based on the number of background subtracted \eflvr events and allowed to fluctuate within its statistical errors.  A random amount of qs-rock background was added according to the value and error of the background estimated from the qs-data at the given \sqdm grid square.

\begin{figure}[htb]
\centerline{\epsfig{file={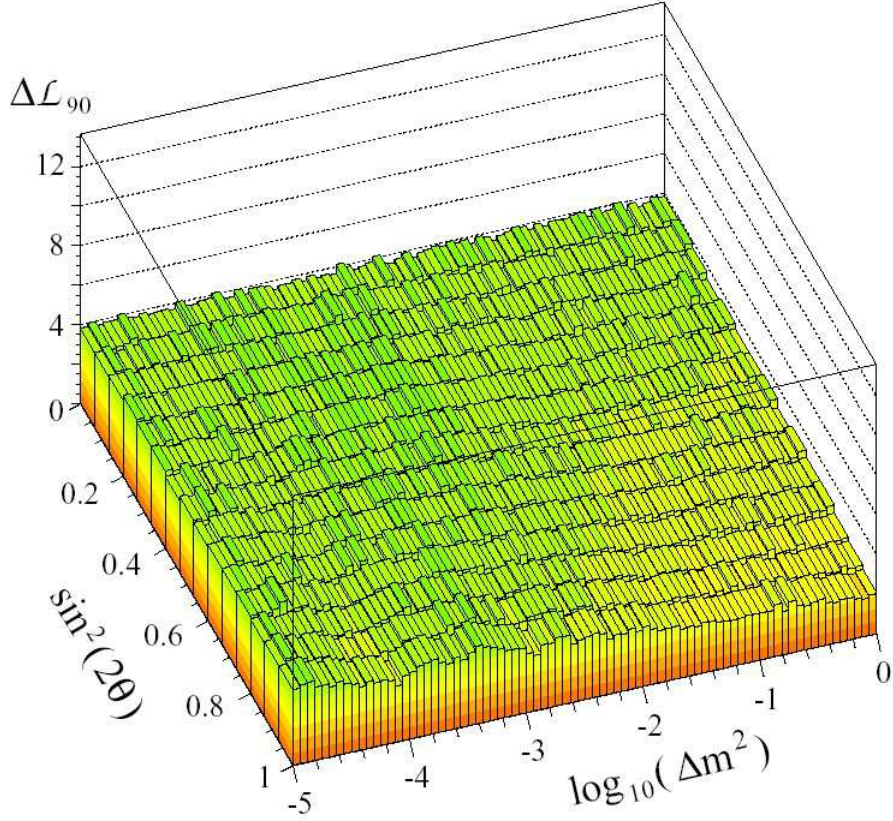} ,width=12.0cm}} 
\caption{The 90\% confidence level surface ($\Delta {\cal L}_{90}$) plotted as a function of \sqsin and log$_{10}(\Delta m^2)$.
}
\label{fig:mc90}
\end{figure}
  
The following systematic effects were incorporated into the analysis:
\begin{enumerate}
\item The energy calibration of the detector has estimated errors of $\pm$7\% on electron showers and $\pm$3\% on muon range. In each MC experiment the calibration was varied within these errors.
\item To allow for the uncertainty in the neutrino flux  as a function of energy, the predicted flux was weighted by a factor $1.0+bE_\nu$ where $b$ was randomly varied from 0.0 for each MC experiment with a Gaussian width of 0.005 ($E_\nu$ in GeV). 
\item The predicted ratio of \nue to \numu events was randomly varied with a Gaussian width of 5\%.
\item To allow for the uncertainties in the neutrino cross-sections, the ratio of quasi-elastic events to inelastic and deep-inelastic events was randomly varied with a Gaussian width of 20\%.
\end{enumerate}
In addition the method automatically included the boundary on \sqsin and the effects of resolution and event misidentification.

 Each MC experiment was analyzed in exactly  the same way as the qs-data, using the same code that produced the results described in Sect.~\ref{oscanal}. The normalization of the MC flux ($A$ parameter) was determined independently for each MC experiment and the fraction of  qs-rock background in each data category was fitted for each MC experiment.

\begin{figure}[htb]
\centerline{\epsfig{file={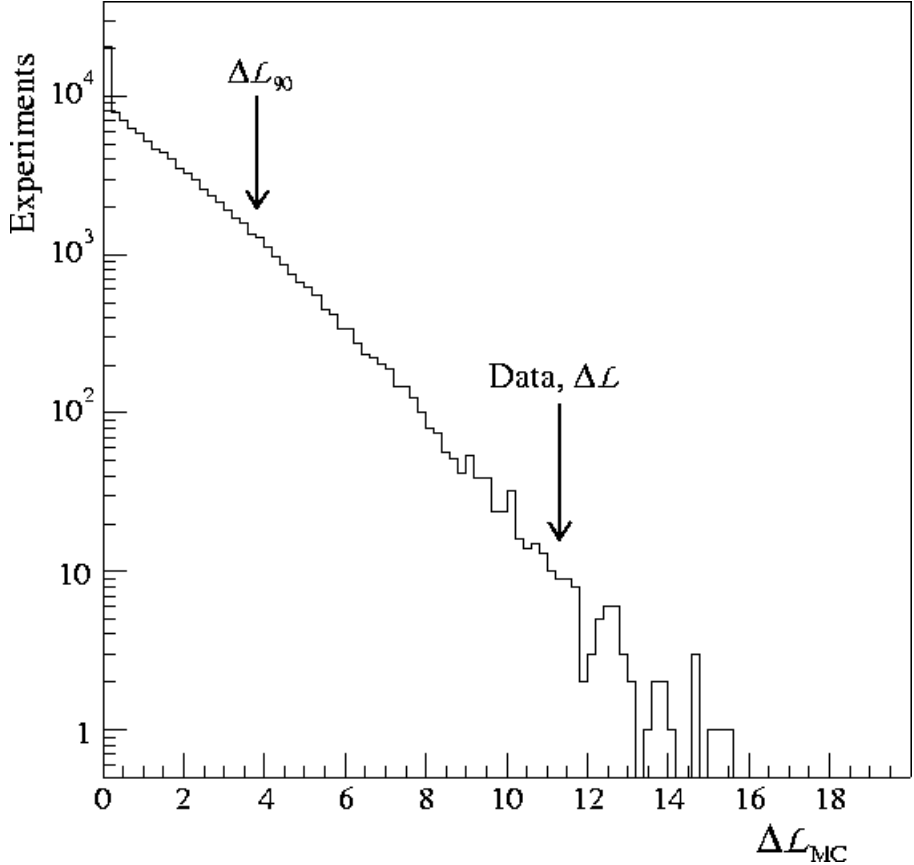} ,width=12.0cm}} 
\caption{The Monte Carlo likelihood distribution $\Delta {\cal L}_{MC}$ for the no oscillation grid square.  The values of the data likelihood ($\Delta {\cal L}$) and $\Delta {\cal L}_{90}$ are shown by the arrows.
}
\label{fig:nolik}
\end{figure}

    One thousand MC experiments were generated at each \sqdm grid square. The best fit grid square in \sqdm was obtained for each experiment, not in general the same as that at which it was generated. The likelihood difference between the generated and best fit \sqdm grid square ($\Delta {\cal L}_{MC}$) was calculated. A histogram of $\Delta {\cal L}_{MC}$ represents the likelihood distribution expected if the truth was at the generated grid square, including the effects of statistics and of the systematic effects. From the histogram the likelihood increase which contains 90\% of the MC experiments is noted ($\Delta {\cal L}_{90}$). The plot of $\Delta {\cal L}_{90}$ as a function of \sqdm is shown in Fig.~\ref{fig:mc90}, defining the 90\% confidence surface. If the data likelihood increase at a given grid square is smaller than $\Delta {\cal L}_{90}$ in that grid square, the square is within the 90\% allowed contour.  Of course other likelihood contours can be obtained by taking different fractions of the MC experiments. A histogram of $\Delta {\cal L}_{MC}$ containing 100,000 experiments generated at the no oscillation grid square of \sqdm is shown in Fig.~\ref{fig:nolik}. The peak at zero is produced by those experiments where the best fit is found at the generated grid square.

\subsection{Confidence region results}
\label{sec:likprob}
 The $\Delta {\cal L}_{90}$ surface in Fig.~\ref{fig:mc90} is higher than 2.3 in the no oscillation region and slopes away from this point.  This is an effect of the inclusion and fitting of the qs-rock background and the constraint that the amount of qs-rock background is positive.  If zero qs-rock events are generated (as frequently happened at these points since the qs-data fit prefers no background) the fit can only produce the same or more qs-rock events.  More qs-rock events require a larger number of events to be lost by oscillation to match the overall normalization, therefore the best fit tends to move to  larger \sqdmns.  The bias in the fits produces a broad likelihood distribution at these points and thus a high $\Delta {\cal L}_{MC}$.  At the opposite corner of the plot, at large \sqdmns, the maximum oscillation signal occurs and no larger decrease in the number of events can be obtained.  Thus  the fit has to remain in this same area of \sqdm  when statistical fluctuations decrease the number of events.  The likelihood distribution is narrow and $\Delta {\cal L}_{MC}$ is decreased.

\begin{figure}[htb]
\centerline{\epsfig{file={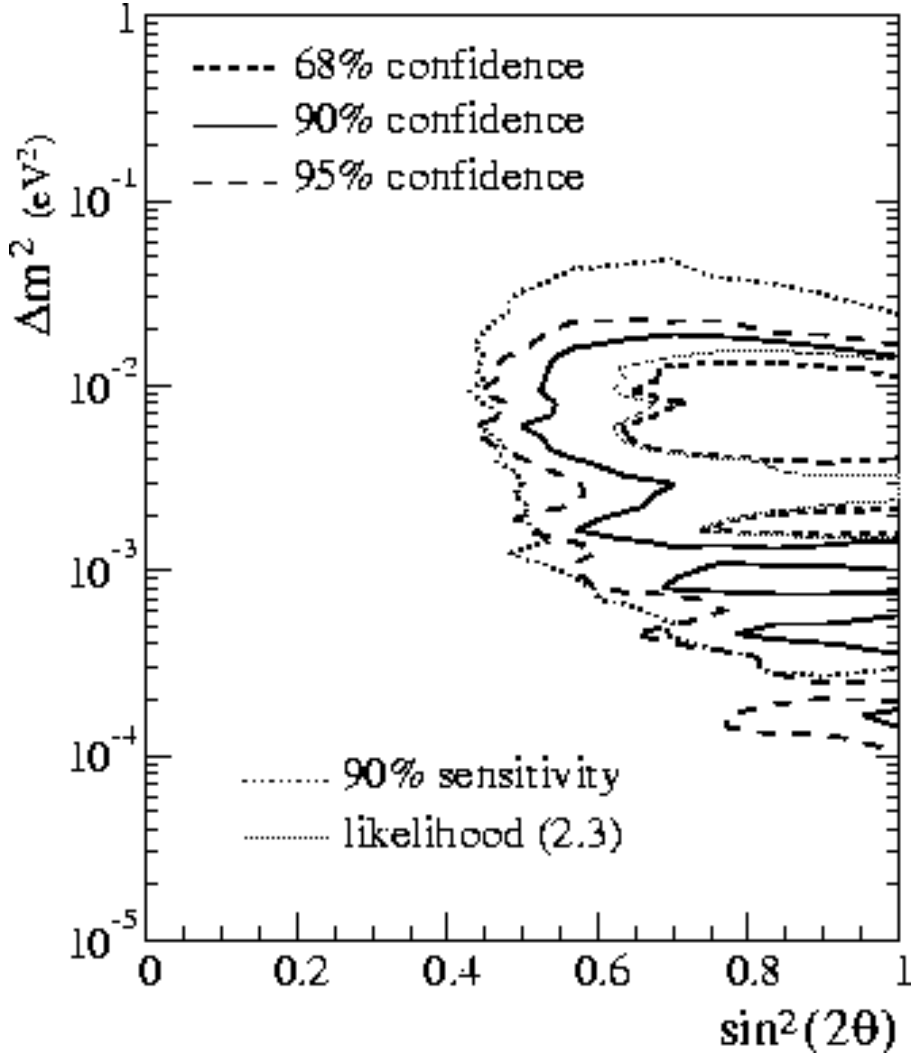} ,width=12.0cm}} 
\caption{ Confidence level contours from the Feldman-Cousins analysis, 68\% (short dashed line), 90\% (thick solid line) and 95\% (long dashed line). The dotted line is the 90\% sensitivity for the best fit \sqdm point.  The thin solid line is the contour defined by a data likelihood rise, $\Delta {\cal L}$, of 2.3.
}
\label{fig:contours}
\end{figure}

Combining  Figs.~\ref{fig:datallh} and \ref{fig:mc90},  a 90\% confidence level contour is obtained and plotted in Fig.~\ref{fig:contours}, together with similar contours for the 68\% and 95\% confidence levels.  At the 68\% ($1\sigma$) level there are two regions. One corresponds to the lower part of the Super-K 90\% confidence region. The other, larger region is at higher \delm and contains the best fit point of this analysis.  The data likelihood is relatively flat in the region immediately below \delmns~=~$10^{-3}$~eV$^2$, which is reflected in the relatively large increase in area going between 90\% and 95\% confidence.

The probability of the no oscillation hypothesis is given by that fraction of the MC experiments at the no oscillation grid square having $\Delta {\cal L}_{MC} > 11.3$, the value of $\Delta {\cal L}$, the data likelihood rise,  for this grid square.  The values of $\Delta {\cal L}_{MC}$ are histogrammed in Fig.~\ref{fig:nolik}.  Fifty-eight of the 100,000 experiments exceeded 11.3, giving a probability for the no oscillation hypothesis of $5.8 \times 10^{-4}$.  This probability takes account of the statistical precision of the experiment and all the systematic effects included in the Feldman-Cousins analysis. 

The dotted line in Fig.~\ref{fig:contours} is the 90\% confidence sensitivity, defined by Feldman and Cousins as the Monte Carlo expectation for the 90\% confidence contour, given this data exposure and the best fit point of the analysis. The data 90\% limit is in reasonable  agreement with the expected sensitivity but lies inside the sensitivity curve, which corresponds closer to the 95\% limit from the qs-data.  This arises because the the flavor ratio for the qs-data is lower than that expected by the MC at the best fit point. 

The thin solid line in Fig.~\ref{fig:contours} is the 90\% confidence region that would be obtained by taking a simple 2.3 rise of the data likelihood, $\Delta {\cal L}$, in Fig.~\ref{fig:datallh}.  The result of the Feldman-Cousins analysis and the inclusion of the systematic errors is to significantly increase the size of the 90\% confidence region.  Roughly the 68\% confidence region of the full analysis corresponds to the 90\% confidence region when these effects are not included.
 
A second, independent Feldman-Cousins based analysis was carried out
on this data \cite{mayly}. There were two main differences from the analysis described in the
previous sections. Firstly, rather than fitting to continuous pdfs of \ltenloe distributions,  binned histograms were used. Fluctuations observed
depending on the bin starting point were resolved by averaging the results from
many different starting points. Secondly, the backgrounds were fixed at the values
determined from depth distributions as described in Sect.~\ref{sec:backsub}. Thus the fits
were made only to the \ltenloe distributions using three parameters,
\sqsinns, \delmns, and the normalization, $A$. Despite the
different procedures, the results of the two analyses were in excellent
agreement.

\section{Flux normalization}
\label{flux}

  The normalization factor $A$ and the amount of qs-rock background are subsidiary outputs from the analysis. They are calculated for each \sqdm grid square. They vary slowly as a function of \sqdmns, both being a minimum at low values of the two  parameters and maximum at high values.  At low values of the parameters there are no oscillation effects predicted, however the qs-data exhibits a large deficit of \muflvr events as described in Sect.~\ref{ratio}.  Thus to get the best fit to the qs-data the overall normalization is reduced and no qs-rock background is added.  Similarly at high values of the parameters the \muflvr events are suppressed by approximately a factor of two at all \loe values. In order to obtain the best compromise between the \eflvr and \muflvr events, the fit includes a larger amount of qs-rock background and requires a larger value of the normalization parameter. Within the 90\% confidence allowed region of \sqdmns, the value of $A$ lies between 85\% and 92\% of the prediction based on the Bartol 96 flux \cite{bartol96} and the NEUGEN neutrino event generator \cite{neugen}.  The number of background qs-rock \muflvr events lies between 5 and 30. These values are in good agreement with those obtained from the fits to the depth distributions alone, described in Sect.~\ref{sec:backsub} and Table~\ref{eventno}.
 
An analysis using the Battistoni 3D atmospheric neutrino flux \cite{Battistoni_3D} together with the neutrino production height prediction of Ref.~\cite{Ruddick} yielded very similar likelihood surfaces but with a best fit $A$ value of 105\%, to be compared with 91\% for the Bartol 96 flux. The major difference between the 3D and 1D fluxes, the peak at low energies towards the horizon, is washed out by the poor experimental resolution in that direction and at those energies. 
     
\section{Conclusions}

\begin{figure}[htb]
\centerline{\epsfig{file={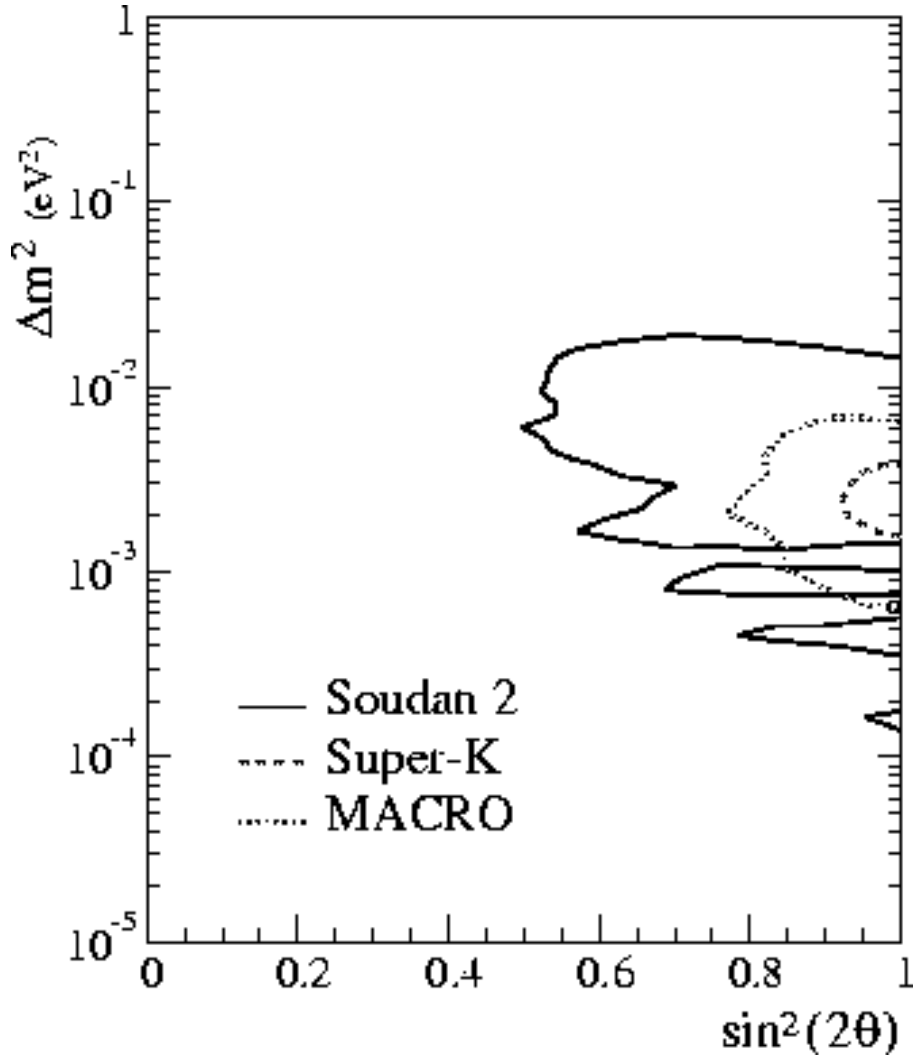} ,width=12.0cm}} 
\caption{ The Soudan 2 90\% confidence allowed region in \sqdm  (solid line) compared with the most recent allowed regions of Super-K (dashed line) \cite{SuperKallowed} and  MACRO (dotted line) \cite{MACROb}.
}
\label{fig:comp}
\end{figure}

The Soudan~2 90\% confidence allowed region is shown in Fig.~\ref{fig:comp} compared with the most recent Super-K \cite{SuperKallowed} and  MACRO \cite{MACROb} allowed regions.  The result presented here is in good agreement  with both experiments.  From the Soudan~2  analysis the probability of the hypothesis of no oscillations is $5.8 \times 10^{-4}$. There is no evidence of a departure from the predicted \loe distribution of the electron events confirming that the oscillation is predominantly \numu to \nutau or \nus. The zenith angle distribution of the $\mu$ flavor  events shows the same depletion as a function of angle as was observed by Super-K. 

 This is the first detailed study of contained and partially contained atmospheric neutrino interactions in an experiment using a  detection technique, an iron calorimeter, which is very different from that of Super-K and previous water Cherenkov detectors. The event detection and reconstruction properties of Soudan~2 are different, and in many cases superior, to those of Super-K but the exposure is much smaller.  The geographical locations and backgrounds of the two experiments are different.  Therefore any detector systematic effect which might simulate neutrino oscillations or bias the determination of oscillation parameters is highly  unlikely to be present in both experiments.  The excellent agreement between the experiments is a strong confirmation of the discovery of neutrino oscillations in the atmospheric neutrino flux.

\section*{Acknowledgments}

   This work was supported by the U.S. Department of Energy, the U.K. Particle 
Physics and Astronomy Research Council, and the State and University of 
Minnesota.  We gratefully acknowledge the Minnesota Department of Natural 
Resources for allowing us to use the facilities of the Soudan Underground 
Mine State Park. We also warmly thank the Soudan 2 mine crew for their dedicated work in building and operating the detector during more than 15 years.

\end{document}